\documentclass[twocolumn,superscriptaddress,showpacs,preprintnumbers,amsmath,amssymb]{revtex4}
\usepackage{graphicx,color}% Include figure files
\usepackage{dcolumn}% Align table columns on decimal point
\usepackage{bm}% bold math
\usepackage{mathrsfs}
\usepackage{amsmath}

\DeclareMathOperator{\im}{Im}

\newcommand{\lc}{\left<}
\newcommand{\rc}{\right>}
\newcommand{\lr}{\left|}
\newcommand{\rl}{\right|}
\newcommand{\lb}{\left(}
\newcommand{\rb}{\right)}
\newcommand{\ls}{\left[}
\newcommand{\rs}{\right]}
\newcommand{\Lb}{\left\{}
\newcommand{\Rb}{\right\}}
\newcommand{\ff}[1]{\frac{1}{#1}}
\newcommand{\scr}[1]{{\mathscr #1}}

\newcommand{\no}{\nonumber\\}

\newcommand{\ppar}[2]{\frac{\partial #1}{\partial #2}}
\newcommand{\mean}[3]{\left<#1\right|#2\left|#3\right>}

\begin{document}
%\begin{CJK*}{GBK}{song}

\title{Feasibility of the finite amplitude method in covariant density functional theory}% Force line breaks with \\

\author{Haozhao Liang}
 \affiliation{RIKEN Nishina Center, Wako 351-0198, Japan}
 \affiliation{State Key Laboratory of Nuclear Physics and Technology, School of Physics,
    Peking University, Beijing 100871, China}

\author{Takashi Nakatsukasa}
 \affiliation{RIKEN Nishina Center, Wako 351-0198, Japan}
 \affiliation{Center for Computational Sciences, University of Tsukuba, Tsukuba 305-8571, Japan}

\author{Zhongming Niu}
 \affiliation{School of Physics and Material Science, Anhui University, Hefei 230039, China}

\author{Jie Meng}
 \affiliation{State Key Laboratory of Nuclear Physics and Technology, School of Physics,
    Peking University, Beijing 100871, China}
 \affiliation{School of Physics and Nuclear Energy Engineering, Beihang University,
    Beijing 100191, China}
 \affiliation{Department of Physics, University of Stellenbosch, Stellenbosch, South Africa}

\date{\today}

\begin{abstract}
Self-consistent relativistic random-phase approximation (RPA) in the radial coordinate representation is established by using the finite amplitude method (FAM).
Taking the isoscalar giant monopole resonance in spherical nuclei as example, the feasibility of the FAM for the covariant density functionals is demonstrated, and the newly developed methods are verified by the conventional RPA calculations.
In the present relativistic RPA calculations, the effects of the Dirac sea can be automatically taken into account in the coordinate-space representation.
The rearrangement terms due to the density-dependent couplings can be implicitly calculated without extra computational costs in both iterative and matrix FAM schemes.
\end{abstract}

\maketitle

%==================Introduction======================================
\section{Introduction}

By reducing the quantum mechanical many-body problems formulated in terms of $N$-body wave functions to the one-body local density distributions, the density functional theory (DFT) of Kohn and Sham \cite{Kohn1965} has accomplished great success in many different fields of modern physics.
No other method achieves comparable accuracy at the same computational costs.
In nuclear physics, the DFT has been widely used since the 1970s \cite{Bender2003}.
In particular, its covariant version in the relativistic framework has received much attention during the past decades.

The covariant density functional theory (CDFT) \cite{Serot1986,Ring1996} takes the Lorentz invariance into account.
In this framework, the representation with large scalar and vector fields, of a few hundred MeV, provides a consistent treatment of the spin degrees of freedom.
The Lorentz symmetry leads to the unification of the time-even and time-odd components in the corresponding functionals.
The Lorentz symmetry also puts stringent restrictions on the number of parameters without reducing the quality of the agreement with experimental data.
Over the years, a large variety of nuclear phenomena have been described successfully by the CDFT \cite{Vretenar2005,Meng2006,Paar2007,Niksic2011}, including the equation of state in symmetric and asymmetric nuclear matter,
ground-state properties of finite spherical and deformed nuclei all over the nuclear chart, collective rotational and vibrational
excitations, fission landscapes, low-lying spectra of transitional nuclei involving quantum phase transitions in
finite nuclear systems, and so on.

Focusing on the vibrational excitations, the random-phase approximation (RPA) \cite{Ring1980} is one of the leading theories applicable to both low-lying excited states and giant resonances.
In the relativistic framework, the self-consistent and quantitative RPA calculations were realized after recognizing the importance of the Dirac sea \cite{Vretenar2000,Ma2001,Ring2001,Kurasawa2003,Haga2005,Wu2006,Kurasawa2013}.
It has been proved that the relativistic RPA is equivalent to the corresponding time-dependent relativistic mean-field (RMF) theory in the small amplitude limit, only if the particle-hole (\textit{ph}) configurations include not only the pairs formed from the occupied and unoccupied Fermi states but also the pairs formed from the Dirac states and occupied Fermi states \cite{Ring2001}.

From then on, great efforts have been dedicated to developing the self-consistent RPA approaches in the relativistic framework \cite{Paar2007}.
The formalism for the nonlinear meson-exchange interactions can be traced back to Refs.~\cite{Ma1997a,Ma1997b}.
For the density-dependent meson-nucleon couplings, the explicit rearrangement terms in the \textit{ph} residual interactions have been derived \cite{Niksic2002}.
The relativistic quasiparticle RPA (QRPA) \cite{Paar2003} has been developed based on the canonical single-nucleon basis of the relativistic Hartree-Bogoliubov theory for giant resonances \cite{Vretenar2003b,Paar2006,Tian2009}, pygmy resonances \cite{Vretenar2004,Paar2005,Paar2009}, and low-lying vibrational states \cite{Ansari2005,Ansari2006,Ansari2007}.
The relativistic (Q)RPA has also been extended to the charge-exchange channels \cite{DeConti1998,Paar2004} for the nuclear spin-isospin resonances \cite{Vretenar2003,Ma2004,Finelli2007,Marketin2012a,Marketin2012b}, $\beta$-decay rates \cite{Niksic2005b,Marketin2007}, muon-capture rates \cite{Marketin2009}, and neutrino-nucleus reactions \cite{Paar2008,Paar2013}.
In addition, the relativistic RPA with finite temperature \cite{Niu2009,Niu2011} and the continuum (Q)RPA \cite{Yang2009,Daoutidis2009,Yang2010,Daoutidis2011} have been established.
To go beyond the mean field, the particle vibrational coupling has also been taken into account \cite{Litvinova2007,Litvinova2010}.

Recently, a fully self-consistent relativistic RPA \cite{Liang2008} has been established based on the relativistic Hartree-Fock theory \cite{Bouyssy1987,Long2006,Long2007}.
It is shown that not only the Gamow-Teller resonances but also the fine structure of spin-dipole resonances can be well reproduced without any readjustment of the energy functional \cite{Liang2008,Liang2012a}.
This self-consistent RPA has also been applied to evaluate the isospin symmetry-breaking corrections to the superallowed $\beta$ transitions for the unitarity test of Cabibbo-Kobayashi-Maskawa matrix \cite{Liang2009}.
The corresponding QRPA \cite{Niu2012arXiv} based on the relativistic Hartree-Fock-Bogoliubov theory \cite{Long2010} has been developed and used for a systematic study of $\beta$-decay half-lives of neutron-rich even-even nuclei with $20 \leqslant Z \leqslant 50$, where the isospin-dependent isoscalar proton-neutron pairing is found to play a very important role.

However, the above investigations are essentially restricted within the spherical symmetry.
The conventional RPA calculations in the matrix form face a big computational challenge when the number of \textit{ph} configurations $N_{ph}$ becomes huge as in the deformed cases.
So far, the only self-consistent deformed (Q)RPA in the relativistic framework was developed by Pe\~na Arteaga \textit{et al.} \cite{PenaArteaga2008,PenaArteaga2009}.
Note that, even in the non-relativistic framework, the deformed (Q)RPA in the matrix form is also a hard task. There are a few recent attempts for the Skyrme energy density functionals in the axially symmetric case \cite{Yoshida2008,Losa2010,Terasaki2010,Yoshida2011,Terasaki2011} and in the triaxial case \cite{Inakura2006},
as well as for the Gogny energy density functionals in the axial case \cite{Peru2008}.
The full three-dimensional calculations have been carried out only using the real-time methods \cite{Nakatsukasa2005,Ebata2010,Stetcu2011,Hashimoto2012}.

As a promising solution for this computational challenge, the so-called finite amplitude method (FAM) was proposed in Ref.~\cite{Nakatsukasa2007}.
In this method, the effects of residual interactions are evaluated in a numerical way by considering a finite density deviation around the ground state.
In such a way, the self-consistent RPA calculations become possible with a little extension of the static Hartree(-Fock) code.
Furthermore, by using the iterative methods for the RPA equation, the computation time is close to a linear dependence on $N_{ph}$, instead of a dependence between $N^2_{ph}$ and $N^3_{ph}$ in the diagonalization scheme \cite{Avogadro2013}.
This advantage is crucial when $N_{ph}$ becomes huge.
In the non-relativistic framework with Skyrme energy density functionals, the feasibility, accuracy, and efficiency of FAM have been demonstrated for the RPA in the three-dimensionally deformed cases in the coordinate-space representation \cite{Nakatsukasa2007,Inakura2009,Inakura2011} and for the QRPA in the spherical \cite{Avogadro2011,Avogadro2013} and axially deformed \cite{Stoitsov2011} cases in the quasiparticle-basis representation.
Iterative algorithms for (Q)RPA solutions have also been developed recently, based on the Arnoldi process \cite{Toivanen2010,Carlsson2012} and on the conjugate gradient method \cite{Imagawa2003}.
The readers are also referred to Ref.~\cite{Nakatsukasa2012} for a recent review.

Therefore, it is worthwhile to develop the self-consistent relativistic RPA by using the finite amplitude method.
In particular, special attentions should be paid to the unique features of covariant density functionals, including the effects of the Dirac sea and the rearrangement terms for the density-dependent interactions.
These rearrangement terms are usually more sophisticated than those in the Skyrme functionals, and cause heavy computations \cite{Niksic2002}.
On the other hand, the covariant density functionals hold the Lorentz invariance, which leads to the unification of their time-even and time-odd components.
This makes the modification in the ground-state code straightforward.

In this work, our premier purpose is to verify the feasibility of the FAM in the CDFT, with special attentions to the Dirac sea and the rearrangement terms.
For a basic demonstration, the self-consistent RPA is established based on the spherical density-dependent point-coupling RMF theory by using the FAM.

The paper is organized as follows:
In Sec.~\ref{sec:TF}, the key formulas of the density-dependent point-coupling RMF theory and the corresponding self-consistent RPA, and the formalism of both iterative and matrix FAM will be presented.
In Sec.~\ref{sec:ND}, the numerical details will be shown with the main focus on the boundary conditions of the $X$ and $Y$ amplitudes in the coordinate-space representation.
In Sec.~\ref{sec:RD}, a benchmark test will be given and the effects of the box size, Dirac sea, and rearrangement terms on the isoscalar giant monopole resonances (ISGMR) will be discussed.
Finally, a summary will be given in Sec.~\ref{sec:SP}.

%
%%==================Theoretical Framework=============================
\section{Theoretical Framework}\label{sec:TF}
\subsection{Point-coupling relativistic mean-field theory}

Successful CDFT can be traced back to the RMF models introduced by Walecka and Serot \cite{Serot1986}.
Since then, the popular RMF models \cite{Ring1996,Vretenar2005,Meng2006} are based on the finite-range meson-exchange representation, in which the nucleus is described as a system of Dirac nucleons that interact with each other via the exchange of mesons.

Recently, the CDFT framework has been reinterpreted by the relativistic Kohn-Sham scheme, and the functionals have been developed based on the zero-range point-coupling interactions \cite{Nikolaus1992}.
In this framework, the meson exchange in each channel is replaced by the corresponding local four-point contact interaction between nucleons.
Such point-coupling model has attracted more and more attentions during the past years due to its simplicity and several other advantages \cite{Burvenich2002,Niksic2008,Zhao2010,Niksic2011,Zhao2011PLB,Zhao2011PRL,Tanimura2012,Zhao2012,Liang2012b,Meng2013}.
In particular, for the present study, by directly expressing the mean-field potentials in terms of nucleon densities and currents, the FAM can be applied in a more straightforward way.

In this section, we recapitulate the key formulas of the point-coupling RMF theory for the FAM calculations, in particular, those related to the currents and space-component of the Coulomb field.

The effective Lagrangian density of the density-dependent point-coupling RMF theory reads \cite{Niksic2008}
\begin{align}
    \mathcal{L} =& \bar\psi (i\gamma^\mu\partial_\mu-M)\psi
        -\ff2\alpha_S(\bar\psi\psi)(\bar\psi\psi)-\ff2\delta_S(\partial_\nu\bar\psi\psi)(\partial^\nu\bar\psi\psi)\no
        &-\ff2\alpha_V(\bar\psi\gamma^\mu\psi)(\bar\psi\gamma_\mu\psi)
        -\ff2\alpha_{tV}(\bar\psi\vec\tau\gamma^\mu\psi)\cdot(\bar\psi\vec\tau\gamma_\mu\psi)\no
        &-e\bar\psi\gamma^\mu A_\mu\frac{(1-\tau_3)}{2}\psi-\frac{1}{4}F^{\mu\nu}F_{\mu\nu},
\end{align}
where $M$ is the nucleon mass, and the field tensor for photons reads $F^{\mu\nu} = \partial^\mu A^\nu - \partial^\nu A^\mu$.
While the coupling parameter $\delta_S$ is a constant, the coupling strengths of the four-nucleon interactions in the scalar ($S$), vector ($V$), and isovector-vector ($tV$) channels are analytical functions with respect to the baryonic density $\rho_b$,
\begin{subequations}\label{eq:alpha}
\begin{align}
    \alpha_S(\rho_b) &= a_S + (b_S + c_Sx)e^{-d_Sx},\\
    \alpha_V(\rho_b) &= a_V + b_V e^{-d_Vx},\\
    \alpha_{tV}(\rho_b) &= b_{tV}e^{-d_{tV}x},
\end{align}
\end{subequations}
with $x=\rho_b/\rho_{\rm sat}$, and $\rho_{\rm sat}$ denotes the saturation density of symmetric nuclear matter.

In this paper, the vectors in coordinate space are denoted by bold type, and vectors in isospin space are denoted by arrows. Greek indices $\mu,\nu$ run over the Minkowski indices $0$, $1$, $2$, and $3$.

The effective Hamiltonian $H$ can be obtained with the general Legendre transformation.
Together with the trial ground state $\lr\Phi_0\rc$ as a Slater determinant, as well as the Hartree and no-sea approximations, the energy functional can be written as
\begin{equation}\label{eq:E}
  E = \lc\Phi_0\rl H\lr\Phi_0\rc = E_k + E_S + E_V + E_{tV} + E_A,
\end{equation}
where the first term is the kinetic energy, and the others correspond to contributions from the scalar, vector, isovector-vector channels and Coulomb field, respectively.

The Dirac equation for nucleons,
\begin{equation}\label{eq:Dirac}
  h\lr\psi_\alpha\rc = \varepsilon_\alpha\lr\psi_\alpha\rc,
\end{equation}
is then obtained by the variation principle.
The one-body mean-field Hamiltonian $h$ is composed of the kinetic term $h_k$, the scalar $h_S$, vector $h_V$, isovector-vector $h_{tV}$, and Coulomb $h_A$ terms, i.e.,
\begin{subequations}
\begin{align}
    h_k &= -i\mathbf{\alpha}\cdot\mathbf{\nabla}+\gamma^0M,\\
    h_S &= \gamma^0(\alpha_S\rho_S + \delta_S\triangle \rho_S),\\
    h_V &= \gamma^0\gamma_\mu\alpha_Vj^\mu_V,\\
    h_{tV} &= \gamma^0\gamma_\mu\alpha_{tV}\vec\tau\cdot\vec j^\mu_{tV},\\
    h_A &= e\gamma^0\gamma_\mu \frac{(1-\tau_3)}{2} A^\mu,
\end{align}
\end{subequations}
together with the additional rearrangement term due to the density-dependent coupling strengths,
\begin{equation}
  h_R = \ff2 \Lb\ppar{\alpha_S}{\rho_b}\rho_S^2 + \ppar{\alpha_V}{\rho_b}j^\mu_Vj_{V\mu} + \ppar{\alpha_{tV}}{\rho_b}\vec j^\mu_{tV}\cdot\vec j_{tV\mu}\Rb.
\end{equation}
In the above expressions, $\rho_S$, $j^\mu_V$, $\vec j^\mu_{tV}$, and $A^\mu$ are the scalar density, the isoscalar and isovector four-currents, as well as the Coulomb field, respectively.
The nuclear baryonic density $\rho_b$ corresponds to the time-component of the isoscalar four-current $j_V^0$.

It is worthwhile to emphasize here that the space-components of the four-currents and the Coulomb field must be kept explicitly for the following applications of FAM, even though they in general vanish in the ground state of systems with the time-reversal symmetry, e.g., even-even nuclei.

For the systems with spherical symmetry, the single-particle wave functions have the form of
\begin{equation}
    \psi_\alpha(\mathbf{r})=\ff r
        \Lb \begin{array}{c}
            iG_a(r) \\ F_a(r)\hat{\sigma}\cdot\hat{\mathbf{r}}
            \end{array} \Rb
            \scr Y_a(\hat{\mathbf{r}})\chi_\ff2(q_a),
\end{equation}
where $\scr Y^{l_a}_{j_am_a}(\hat{\mathbf{r}})$ are the spherical harmonics spinors, $\chi_\ff2(q_a)$ the isospinors.
The single-particle eigenstates are specified by the set of quantum numbers $\alpha = (a,m_a) = (q_a, n_a, l_a, j_a, m_a)$, and the good quantum number $\kappa_a = \mp(j_a + 1/2)$ for $j_a = l_a \pm 1/2$.
Within this phase convention between the upper and lower components, the wave functions $G(r)$ and $F(r)$ can be simultaneously chosen as real functions for the ground-state descriptions.
In contrast, for the FAM built beyond, both $G(r)$ and $F(r)$ become complex functions, so one should be careful to distinguish them from their complex conjugates $G^*(r)$ and $F^*(r)$ from the very beginning.

The radial Dirac equation reads
\begin{widetext}
\begin{equation}\label{eq:rDirac}
    \lb \begin{array}{cc}
        M+\Sigma_S(r)+\Sigma_0(r) & -\frac{d}{dr}+\frac{\kappa_a}{r}+\Sigma_V(r) \\
        \frac{d}{dr}+\frac{\kappa_a}{r}-\Sigma_V(r) & -M-\Sigma_S(r)+\Sigma_0(r)
    \end{array} \rb
    \lb \begin{array}{c}
        G_a(r) \\ F_a(r)
    \end{array} \rb
    =\varepsilon_a
    \lb \begin{array}{c}
        G_a(r) \\ F_a(r)
    \end{array} \rb
\end{equation}
with the scalar and vector potentials
\begin{subequations}\label{eq:Sigma}
\begin{align}
    \Sigma_S(r) &= \alpha_S \rho_S(r) + \delta_S\lb\rho''_S(r)+\frac{2}{r}\rho'_S(r)\rb,\\
    \Sigma_0(r) &= \alpha_V \rho_V(r) + \alpha_{tV}\rho_{tV}(r)\tau_3 +e\frac{1-\tau_3}{2}A_0(r) + \Sigma_R(r),\\
    \Sigma_V(r) &= \alpha_V j_V(r) + \alpha_{tV}j_{tV}(r)\tau_3 +e\frac{1-\tau_3}{2}A_V(r).
\end{align}
\end{subequations}
The rearrangement terms only contribute to the time-component of the vector potential, which read
\begin{equation}\label{eq:SigmaR}
    \Sigma_R(r)=\ff2\Lb \ppar{\alpha_S}{\rho_b}\rho_S^2(r) + \ppar{\alpha_V}{\rho_b}(\rho_V^2(r)+j_V^2(r))
        + \ppar{\alpha_{tV}}{\rho_b}(\rho_{tV}^2(r)+j_{tV}^2(r))\Rb.
\end{equation}
\end{widetext}
The densities and currents are expressed as
\begin{subequations}\label{eq:dens}
\begin{align}
    \rho_S^{(q_a)}&=
        \ff{4\pi r^2}\sum^{q_a}\hat j_a^2
        \ls G^*_a(r)G_a(r)-F^*_a(r)F_a(r)\rs,\\
    \rho_V^{(q_a)}&=
        \ff{4\pi r^2}\sum^{q_a}\hat j_a^2
        \ls G^*_a(r)G_a(r)+F^*_a(r)F_a(r)\rs,\\
    j_V^{(q_a)}&=
        \ff{4\pi r^2}\sum^{q_a}\hat j_a^2
        \ls G^*_a(r)F_a(r)-F^*_a(r)G_a(r)\rs,
\end{align}
\end{subequations}
with $\hat j^2_a = 2j_a+1$.
The isoscalar densities and currents are the sum of the neutron and proton contributions, while the isovector ones are the differences between the neutron and proton contributions.
The Coulomb fields are calculated with the Green's function method, i.e.,
\begin{subequations}\label{eq:Coulomb}
\begin{align}
    A_0(r) &= e\int dr' {r'}^2 \rho_V^{(p)}(r') \ff{r_>},\\
    A_V(r) &= \frac{e}{3}\int dr' {r'}^2 j_V^{(p)}(r') \frac{r_<}{r^2_>},
\end{align}
\end{subequations}
where $r_>\equiv\max\{r,r'\}$ and $r_<\equiv\min\{r,r'\}$.

\subsection{Linear response and random-phase approximation}

The RPA equation is known to be equivalent to the time-dependent Hartree(-Fock) equation in the small amplitude limit \cite{Ring1980}.
In order to make the FAM clear in the next section, we first briefly recall the derivation of the standard RPA equation by following the notations in Ref.~\cite{Nakatsukasa2007}.

The static Hartree or Hartree-Fock equation,
\begin{equation}
    [h[\rho], \rho] = 0,
\end{equation}
determines the ground-state density $\rho=\rho_0$ satisfying $\rho^2 = \rho$, and the one-body mean-field Hamiltonian $h_0 = h[\rho_0]$.

When a time-dependent external perturbation $V_{\rm ext}(t)$ is present, the density deviation $\delta\rho(t)\equiv\rho(t)-\rho_0$ obeys \begin{equation}
  i\frac{d}{dt}\delta\rho(t)=[h_0,\delta\rho(t)]+[\delta h(t)+V_{\rm ext}(t),\rho_0]
\end{equation}
as a linear response to the weak perturbation.
In the frequency representation, the above equation is expressed as
\begin{equation}\label{eq:linear}
    \omega\delta\rho(\omega)=[h_0,\delta\rho(\omega)]+[\delta h(\omega)+V_{\rm ext}(\omega),\rho_0].
\end{equation}

In practical calculations, it is convenient to adopt the single-particle (Kohn-Sham) orbitals to represent the density matrix,
\begin{equation}
    \rho(t)=\sum_{i=1}^A \lr\psi_i(t)\rc\lc\psi_i(t)\rl.
\end{equation}
As a result, the density deviation in the frequency representation can be expressed as
\begin{equation}
  \delta\rho(\omega) = \sum_{i=1}^A \{\lr X_i(\omega)\rc\lc\phi_i\rl+\lr\phi_i\rc\lc Y_i(\omega)\rl\}
\end{equation}
with the so-called forward $X(\omega)$ and backward $Y(\omega)$ amplitudes and the occupied eigenstates $\{\lr\phi_i\rc\}$ of $h_0$ in Eq.~(\ref{eq:Dirac}).
It is slightly tricky that one must take the ket $\lr X_i(\omega)\rc$ and bra $\lc Y_i(\omega)\rl$ states independent, since $\delta\rho(\omega)$ is not Hermitian.
But, this point is in fact well known as the solutions of the RPA equation shown below.
Hereafter, $\lr\phi_a\rc$ represent the eigenstates of $h_0$, and indices $i,j$ ($m,n$) run over the hole (particle) states.

By expanding the $X(\omega)$ and $Y(\omega)$ amplitudes on the basis of particle states,
\begin{subequations}\label{eq:XYmi}
\begin{align}
    \lr X_i(\omega)\rc&=\sum_{m>A}\lr\phi_m\rc X_{mi}(\omega),\\
    \lr Y_i(\omega)\rc&=\sum_{m>A}\lr\phi_m\rc Y^*_{mi}(\omega),
\end{align}
\end{subequations}
one can derive the well-known RPA equation in the matrix form,
\begin{widetext}
\begin{equation}\label{eq:RPA}
    \Lb\lb\begin{array}{cc}
      \mathcal{A}_{mi,nj} & \mathcal{B}_{mi,nj} \\
      \mathcal{B}^*_{mi,nj} & \mathcal{A}^*_{mi,nj}
    \end{array}\rb
    -\omega
    \lb\begin{array}{cc}
      1 & 0 \\
      0 & -1
    \end{array}\rb\Rb
    \lb\begin{array}{c}
      X_{nj}(\omega) \\
      Y_{nj}(\omega)
    \end{array}\rb
    =-\lb\begin{array}{c}
      f_{mi}(\omega) \\
      g_{mi}(\omega)
    \end{array}\rb.
\end{equation}
The RPA matrices $\mathcal{A}$ and $\mathcal{B}$ and vectors $\vec f$ and $\vec g$ read
\begin{subequations}\label{eq:AB}
\begin{align}
    \mathcal{A}_{mi,nj}
        &=(\epsilon_m-\epsilon_i)\delta_{mn}\delta_{ij}+\mean{\phi_m}{\left.\frac{\partial h}{\partial\rho_{nj}}\right|_{\rho=\rho_0}}{\phi_i}
        =(\epsilon_m-\epsilon_i)\delta_{mn}\delta_{ij}+\mean{\phi_m\phi_j}{V_{ph}}{\phi_n\phi_i},\\
    \mathcal{B}_{mi,nj}
        &=\mean{\phi_m}{\left.\frac{\partial h}{\partial\rho_{jn}}\right|_{\rho=\rho_0}}{\phi_i}
         =\mean{\phi_m\phi_n}{V_{ph}}{\phi_j\phi_i},\\
    f_{mi}&=\mean{\phi_m}{V_{\rm ext}(\omega)}{\phi_i},\qquad
    g_{mi}=\mean{\phi_i}{V_{\rm ext}(\omega)}{\phi_m}.
\end{align}
\end{subequations}

For the self-consistent RPA calculations \cite{Ring1980}, the particle-hole residual interactions $V_{ph}$ should be strictly derived from the second derivative of the energy functional $E$ shown in Eq.~(\ref{eq:E}).
The \textit{ph} residual interactions for the point-coupling RMF theory with nonlinear couplings can be found in Ref.~\cite{Niksic2005}.
In contrast,  the density dependence in the coupling strengths $\alpha$ introduces additional rearrangement terms in $V_{ph}$ \cite{Niksic2002}.
Explicitly, the \textit{ph} residual interactions are composed of
\begin{subequations}\label{eq:Vph}
\begin{align}
          V^S_{ph}(1,2) &= \Lb\alpha_S [\gamma^0]_1[\gamma^0]_2
          + \frac{\partial\alpha_S}{\partial\rho_b}\rho_S([\gamma^0]_1[\mathbb{I}]_2+[\mathbb{I}]_1[\gamma^0]_2)
          + \ff2\frac{\partial^2\alpha_S}{\partial\rho_b^2}\rho^2_S[\mathbb{I}]_1[\mathbb{I}]_2-\delta_S [\gamma^0\mathbf{\nabla}]_1\cdot[\gamma^0\mathbf{\nabla}]_2\Rb\delta(\mathbf{r}_1-\mathbf{r}_2),\\
          V^V_{ph}(1,2) &= \Lb\alpha_V [\gamma^0\gamma^\mu]_1[\gamma^0\gamma_\mu]_2
          + 2\frac{\partial\alpha_V}{\partial\rho_b}\rho_V[\mathbb{I}]_1[\mathbb{I}]_2
          + \ff2\frac{\partial^2\alpha_V}{\partial\rho_b^2}\rho^2_V[\mathbb{I}]_1[\mathbb{I}]_2\Rb
          \delta(\mathbf{r}_1-\mathbf{r}_2),\\
          V^{tV}_{ph}(1,2) &= \Lb\alpha_{tV} [\gamma^0\gamma^\mu\vec\tau]_1\cdot[\gamma^0\gamma_\mu\vec\tau]_2
          + \frac{\partial\alpha_{tV}}{\partial\rho_b}\rho_{tV}([\tau_3]_1[\mathbb{I}]_2+[\mathbb{I}]_1[\tau_3]_2)
          + \ff2\frac{\partial^2\alpha_{tV}}{\partial\rho_b^2}\rho^2_{tV}[\mathbb{I}]_1[\mathbb{I}]_2\Rb
          \delta(\mathbf{r}_1-\mathbf{r}_2),\\
          V^{A}_{ph}(1,2) &= \frac{e^2}{4\pi}[\gamma^0\gamma^\mu\frac{1-\tau_3}{2}]_1[\gamma^0\gamma_\mu\frac{1-\tau_3}{2}]_2
            \ff{|\mathbf{r}_1-\mathbf{r}_2|},
\end{align}
\end{subequations}
\end{widetext}
where $\mathbb{I}$ denotes the $4\times 4$ unit matrix.
The rearrangement terms correspond to those containing $\partial\alpha/\partial\rho_b$ or $\partial^2\alpha/\partial\rho_b^2$.
They are calculated term by term separately in the conventional RPA calculations.

Meanwhile, it is also important to emphasize the effects of the Dirac sea.
The relativistic RPA is equivalent to the time-dependent RMF theory in the small amplitude limit, only when the particle states $m,n$ include not only the states above the Fermi surface but
also the states in the Dirac sea \cite{Ring2001}.
It is due to the no-sea approximation used in the ground-state calculations.
In other words, the ensemble of all these unoccupied states together provides a complete set of basis for particle states.

\subsection{Iterative finite amplitude method}\label{sec:iFAM}

In Ref.~\cite{Nakatsukasa2007}, the FAM was proposed as a simpler and more efficient approach to the solutions of the linear response equation~(\ref{eq:linear}).
This method does not require explicit evaluation of the residual interactions $\delta h/\delta\rho$ as in Eq.~(\ref{eq:AB}).
Instead, by multiplying with the ket $\lr\phi_i\rc$ and bra $\lc\phi_i\rl$ of only hole states on both sides of Eq.~(\ref{eq:linear}), respectively, one has
\begin{subequations}\label{eq:FAM}
\begin{align}
    \omega\lr X_i(\omega)\rc
        &= (h_0-\epsilon_i)\lr X_i(\omega)\rc+\hat Q(V_{\rm ext}(\omega)+\delta h(\omega))\lr\phi_i\rc,\\
    \omega^*\lr Y_i(\omega)\rc
        &= -(h_0-\epsilon_i)\lr Y_i(\omega)\rc-\hat Q(V^\dag_{\rm ext}(\omega)+\delta h^\dag(\omega))\lr\phi_i\rc,
\end{align}
\end{subequations}
where $\hat Q=1-\sum_j\lr\phi_j\rc\lc\phi_j\rl$ is a projection operator onto the particle space.

The induced fields $\delta h(\omega)$ and $\delta h^\dag(\omega)$ shown above are calculated by using the following finite difference with a sufficiently small number $\eta$:
\begin{equation}\label{eq:dh}
  \delta h(\omega)=\ff\eta(h[\lc\psi'\rl,\lr\psi\rc]-h[\lc\phi\rl,\lr\phi\rc])
\end{equation}
with $\lc\psi'_i\rl=\lc\phi_i\rl+\eta\lc Y_i(\omega)\rl$ and $\lr\psi_i\rc=\lr\phi_i\rc+\eta\lr X_i(\omega)\rc$, and
\begin{equation}\label{eq:dh+}
  \delta h^\dag(\omega)=\ff\eta(h[\lc\psi'\rl,\lr\psi\rc]-h[\lc\phi\rl,\lr\phi\rc])
\end{equation}
with
$\lc\psi'_i\rl=\lc\phi_i\rl+\eta\lc X_i(\omega)\rl$ and $\lr\psi_i\rc=\lr\phi_i\rc+\eta\lr Y_i(\omega)\rc$.

For the present calculations with spherical symmetry, it is convenient to rewrite Eq.~(\ref{eq:FAM}) in coordinate space.
Assuming the monopole perturbation,
\begin{subequations}
\begin{align}
    V_{\rm ext}(\mathbf{r},\omega) &= V_{\rm ext}(r, \omega)Y_{00}(\hat{\mathbf{r}}),\\
    X_i(\mathbf{r}) &= X_i(r)Y_{00}(\hat{\mathbf{r}}),\\
    Y_i(\mathbf{r}) &= Y_i(r)Y_{00}(\hat{\mathbf{r}}),
\end{align}
\end{subequations}
the corresponding radial FAM equations read
\begin{subequations}\label{eq:iFAM}
\begin{align}
    &\hat Q\ls(h_0(r)-\epsilon_i-\omega)X_i(r,\omega) +\delta h(r,\omega)\phi_i(r)\rs\no
    =&  -\hat Q V_{\rm ext}(r,\omega)\phi_i(r),\\
    &\hat Q\ls(h_0(r)-\epsilon_i+\omega^*)Y_i(r,\omega) +\delta h^\dag(r,\omega)\phi_i(r)\rs^*\no
    =& -\hat Q \ls V^\dag_{\rm ext}(r,\omega)\phi_i(r)\rs^*.
\end{align}
\end{subequations}
In the relativistic framework, $h(r)$ is a $2\times2$ matrix as shown in the radial Dirac equation (\ref{eq:rDirac}), and $\phi_i(r)=(G_i(r)~F_i(r))^T$. Therefore, the $X$ and $Y$ amplitudes are also composed of the upper and lower components,
\begin{equation}\label{eq:XYr}
  X_i(r) =
  \lb\begin{array}{c}
      X_{Gi}(r) \\
      X_{Fi}(r)
    \end{array}\rb,\quad
  Y_i(r) =
  \lb\begin{array}{c}
      Y_{Gi}(r) \\
      Y_{Fi}(r)
    \end{array}\rb.
\end{equation}

As emphasized in the previous section, the effects of the Dirac sea must be taken into account, which is expressed in an explicit way in the conventional expansions (\ref{eq:XYmi}).
In contrast, here the $X$ and $Y$ amplitudes are expanded on the mesh points $\{r_k\}$ in coordinate space.
In such a way, on one hand, the effects of the Dirac sea cannot be identified or isolated; on the other hand, from the mathematical point of view, the coordinate space $\sum_{\mathbf{r}}\lr\mathbf{r}\rc\lc\mathbf{r}\rl-\sum_j\lr\phi_j\rc\lc\phi_j\rl$, can also provide a complete set of basis for particle states.

The induced fields $\delta h(r)$ and $\delta h^\dag(r)$ are evaluated by using Eqs.~(\ref{eq:dh}) and (\ref{eq:dh+}).
The procedure in practice is as follows: with a given set of $\{X_i(r)\}$ and $\{Y_i(r)\}$, one calculates the nucleon densities and currents [Eq.~(\ref{eq:dens})], new coupling strengths [Eq.~(\ref{eq:alpha})], Coulomb fields [Eq.~(\ref{eq:Coulomb})], rearrangement self-energy [Eq.~(\ref{eq:SigmaR})], scalar and vector potentials [Eq.~(\ref{eq:Sigma})], and then the one-body Hamiltonian $h(r)$ [Eq.~(\ref{eq:rDirac})], sequentially.
Since now the $X(r)$ and $Y(r)$ amplitudes are independent due to the non-Hermitian nature of $\delta h(r)$ and $\delta h^\dag(r)$, it is clear that the nucleon currents are no longer vanishing.
This is the reason why these time-odd terms must be kept from the beginning.

In order to include both the normal and rearrangement terms in the \textit{ph} residual interactions as explicitly shown in Eq.~(\ref{eq:Vph}), one simply needs to re-calculate the coupling strengths $\alpha$ appearing in Eq.~(\ref{eq:Sigma}) and their derivatives $\partial\alpha/\partial\rho_b$ in Eq.~(\ref{eq:SigmaR}) by using Eq.~(\ref{eq:alpha}) for each given set of $\{X_i(r)\}$ and $\{Y_i(r)\}$.
If one skips this step, i.e., keeps $\alpha$ and $\partial\alpha/\partial\rho_b$ always unchanged, the consequence is that the normal terms in $V_{ph}$ remain, but all of the rearrangement terms are neglected.

This FAM equation is a standard linear algebraic equation of the form, $\mathcal{A}\vec x=\vec b$, which can be solved within the iterative scheme.
In such a way, we do not need to construct the matrix elements of $\mathcal{A}$ explicitly, but only to evaluate $\mathcal{A}\vec x$ for a given vector $\vec x$.
In the following, we denote this iterative finite amplitude method as i-FAM.

Adopting the $\omega$-independent local external field $V_{\rm ext}(r,\omega)= O(r)$,
the corresponding transition strengths can be calculated with the solutions of Eq.~(\ref{eq:iFAM}) as
\begin{align}
  \frac{d B(\omega;O)}{d\omega}
  \equiv&\sum_n |\mean{\Phi_n}{O}{\Phi_0}|^2\delta(\omega-E_n)\no
  =& -\ff\pi \im \sum_{i}\hat j_i^2\int dr\{\phi^\dag_i(r)O^\dag(r)X_i(r,\omega)\no
    &+Y^\dag_i(r,\omega)O^\dag(r)\phi_i(r)\}.
\end{align}

\subsection{Matrix finite amplitude method}\label{sec:mFAM}

We introduce another usage of FAM, the so-called matrix finite amplitude method (m-FAM) shown in Ref.~\cite{Avogadro2013}.
In this method, the RPA matrices $\mathcal{A}$ and $\mathcal{B}$ are explicitly constructed, but the tedious calculations concerning the \textit{ph} residual interactions $V_{ph}$ in Eqs.~(\ref{eq:AB}) and (\ref{eq:Vph}) can be avoided.

First of all, both the occupied and unoccupied eigenstates of $h_0$, $\{\lr\phi_i\rc\}$ and $\{\lr\phi_m\rc\}$, are calculated at the ground state.
Then, instead of dealing with $V_{ph}$, the kernels $\partial h/\partial\rho$ in Eq.~(\ref{eq:AB}) are directly calculated with finite difference provided the real parameter $\eta$ is small enough to neglect the higher-order terms, i.e.,
\begin{equation}
  \left.\frac{\partial h}{\partial\rho_{nj}}\right|_{\rho=\rho_0} =
  \ff\eta(h[\lc\psi'\rl,\lr\psi\rc]-h[\lc\phi\rl,\lr\phi\rc]).
\end{equation}
The key point here is to keep all $\lc\psi'_i\rl=\lc\phi_i\rl$ and $\lr\psi_i\rc=\lr\phi_i\rc$ unchanged, except for the specific orbital $j$ which slightly mixes with another specific orbital $n$ as $\lr\psi_j\rc=\lr\phi_j\rc+\eta\lr \phi_n\rc$.
In the same way,
\begin{equation}
  \left.\frac{\partial h}{\partial\rho_{jn}}\right|_{\rho=\rho_0} =
  \ff\eta(h[\lc\psi'\rl,\lr\psi\rc]-h[\lc\phi\rl,\lr\phi\rc]),
\end{equation}
by keeping all $\lc\psi'_i\rl=\lc\phi_i\rl$ and $\lr\psi_i\rc=\lr\phi_i\rc$ unchanged, but slightly mixing specific orbitals $j$ with $n$ as $\lc\psi'_j\rl=\lc\phi_j\rl+\eta\lc \phi_n\rl$.

To include the effects of the Dirac sea, states $n$ run over the unoccupied states in both Fermi and Dirac sea.
To include the effects of the rearrangement terms, one follows the same procedure as that in i-FAM shown above.

%
%%==================Numerical Details===============================================
\section{Numerical Details}\label{sec:ND}

For all the calculations in this paper, the density-dependent point-coupling RMF parametrization DD-PC1 \cite{Niksic2008} is used and the spherical symmetry is assumed.
For the ground-state calculations, the radial Dirac equation (\ref{eq:rDirac}) is solved in coordinate space by the fourth-order Runge-Kutta method, also known as the shooting method, within a spherical box with a box radius $R$ and a mesh size $dr$ \cite{Meng1998}.
The mesh size is fixed as $dr=0.1$~fm, while the choice of box size $R$ will be discussed below.

For the conventional RPA and m-FAM calculations, the single-particle energy truncation for constructing the RPA matrices $\mathcal{A}$ and $\mathcal{B}$ in Eq.~(\ref{eq:RPA}) is $[-M, M+200~\mbox{MeV}]$, i.e., all the bound states in the Dirac sea are taken into account.
As an example, the corresponding number of \textit{ph} configurations $N_{ph}$ for $J^\pi=0^+$ excitations in $^{208}$Pb is $1355$ with $R=25$~fm, where $524$ of them are formed with the particle states in the Dirac sea.
The convergency of this truncation has been examined.
The subroutine \verb"rg.f" in EISPACK library is used to diagonalize the non-symmetric real RPA matrix.
In m-FAM, the parameter $\eta$ is taken as $\eta=10^{-2}$.

For the i-FAM calculations, the frequency $\omega= E+i\Gamma/2$ contains an imaginary part, and the corresponding Lorentzian smearing parameter is $\Gamma=1$~MeV.
The first derivative of $\{X_i(r)\}$ and $\{Y_i(r)\}$ with respective to $r$ is performed by the nine-point formula with the boundary conditions discussed below.
The parameter $\eta$ differs for every iteration to ensure the linearity \cite{Nakatsukasa2007,Inakura2009}:
\begin{equation}
  \eta=\frac{10^{-6}}{\max\{N(X),N(Y)\}},\quad
  N(\psi) = \frac{1}{A}\sqrt{\sum_{i=1}^A\lc\psi_i|\psi_i\rc}.
\end{equation}
The truncated version of generalized conjugate residual (GCR) method \cite{Saad2003}, also called ORTHOMIN method, is used as the iterative solver, where at maximum 1000 iterations are stored.
The convergent criterion is $||\mathcal{A}\vec x-\vec b||^2/||\vec b||^2<10^{-6}$, which provides the relative accuracy $10^{-3}$ for the transition strengths.

\subsection{Boundary conditions}

Before further discussions, it is worthwhile to examine the boundary conditions of the $X$ and $Y$ amplitudes (\ref{eq:XYr}) in the coordinate-space representation.
It turns out somehow tricky since these amplitudes contain two components instead of one as in the non-relativistic case.

The boundary conditions for the ground-state radial Dirac equation (\ref{eq:rDirac}) used in the shooting method are following \cite{Meng1998}:
(i) At the origin point, $G(r)|_{r=0}=F(r)|_{r=0}=0$.
(ii) At small distance $r\rightarrow0$, $G(r)\propto r^{-\kappa}, F(r)\propto r^{1-\kappa}$ for $\kappa<0$ and $G(r)\propto r^{1+\kappa}, F(r)\propto r^{\kappa}$ for $\kappa>0$.
(iii) At the box boundary, $G(r)|_{r=R}=0$, but $F(r)|_{r=R}$ must have a non-vanishing value, otherwise the whole wave function will be identically zero.
The value of $F(r)|_{r=R}$ is eventually determined by the normalization condition.

Accordingly, the consistent boundary conditions of $X(r)$ and $Y(r)$ used in i-FAM include:
(i) At the origin point, $X_G(r)|_{r=0} = X_F(r)|_{r=0} = Y_G(r)|_{r=0} = Y_F(r)|_{r=0} = 0$.
(ii) At small distance $r\rightarrow0$, $X_G(r), Y_G(r)$ are odd functions and $X_F(r), Y_F(r)$ are even functions for even $l$, while $X_G(r), Y_G(r)$ are even functions and $X_F(r), Y_F(r)$ are odd functions for odd $l$.
(iii) The remaining but critical point is the boundary conditions at the box boundary $r=R$.
In addition, outside the box, $X_G(r)|_{r>R} = X_F(r)|_{r>R} = Y_G(r)|_{r>R} = Y_F(r)|_{r>R} = 0$, since it is an area out of consideration.

\begin{figure}
\centering
\includegraphics[width=8cm]{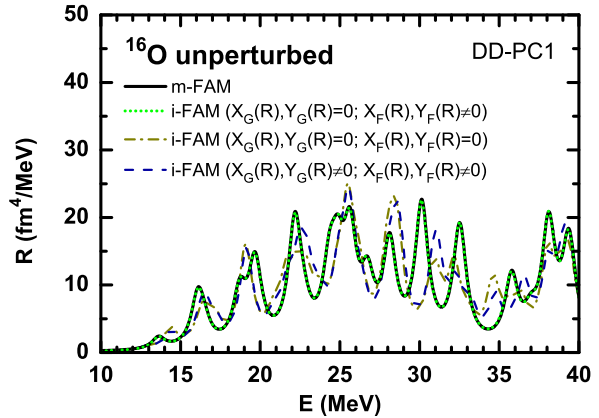}
\caption{(Color online) The $J^\pi = 0^+$ unperturbed excitation strengths in $^{16}$O calculated by the matrix finite amplitude method (m-FAM) (solid line) with $R=20$~fm and $dr=0.1$~fm.
The corresponding results calculated by the iterative finite amplitude method (i-FAM) with different box boundary conditions are shown with the short-dotted, dash-dotted, and dashed lines, respectively.
A Lorentzian smearing parameter $\Gamma = 1$~MeV is used.
    \label{Fig1}}
\end{figure}

In order to verify the boundary conditions at $r=R$, in Fig.~\ref{Fig1}, we show with the solid line the $J^\pi = 0^+$ unperturbed excitation strengths in $^{16}$O calculated by m-FAM with a box radius $R=20$~fm and a mesh size $dr=0.1$~fm.
In m-FAM, the particle states $\{\phi_m(r)\}$ correspond to the eigenstates of $h_0(r)$ with the boundary conditions used in the shooting method.
Naturally, these boundary conditions are consistent with the ground-state description.

In the same figure, the corresponding results calculated by i-FAM with different boundary conditions of $X(r)$ and $Y(r)$ at $r=R$ are shown for comparison.
The results obtained by constraining $X_G(r)|_{r=R} = X_F(r)|_{r=R} = Y_G(r)|_{r=R} = Y_F(r)|_{r=R} = 0$ are shown with the dash-dotted line, those obtained by constraining only $X_G(r)|_{r=R} = Y_G(r)|_{r=R} = 0$ are shown with the short-dotted line, and those obtained without any constraint at $r=R$ are shown with the dashed line.
The tiny difference between the dash-dotted and dashed lines shows the effect of changing the box size by one mesh point $dr$,
but the visible difference between the short-dotted line and the other two is due to the different prescriptions for the upper and lower components at the same position.
It can be clearly seen that only the short-dotted one with proper boundary conditions is identical to the m-FAM result.

Therefore, the consistent boundary conditions around $r=R$ for the $X(r)$ and $Y(r)$ amplitudes in the i-FAM calculations read
\begin{equation}
    X_G(r)|_{r\geqslant R} = X_F(r)|_{r>R} = Y_G(r)|_{r\geqslant R} = Y_F(r)|_{r>R} = 0.
\end{equation}

%
%%==================Results and discussion===============================================
\section{Results and Discussion}\label{sec:RD}

In the following discussions, we take the stable and radioactive neutron-rich doubly magic nuclei, $^{208}$Pb and $^{132}$Sn, as examples.
It has been shown that the RMF theory can in general nicely reproduce the corresponding ground-state properties (see e.g., Ref.~\cite{Liang2011}).

\subsection{Benchmark tests}

\begin{figure}
\centering
\includegraphics[width=8cm]{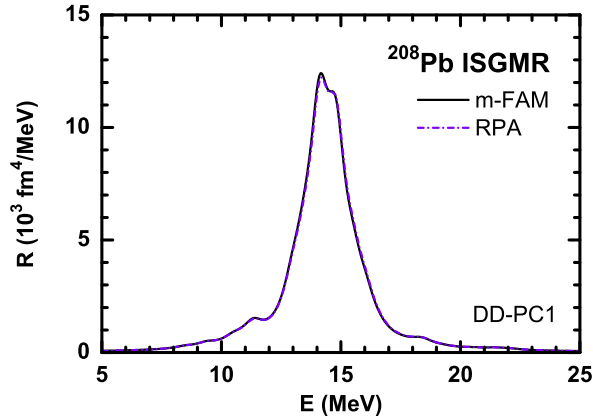}
\caption{(Color online) Isoscalar giant monopole resonance (ISGMR) in $^{208}$Pb calculated by m-FAM (solid line) and conventional RPA (short-dash-dotted line).
A Lorentzian smearing parameter $\Gamma = 1$~MeV is used.
    \label{Fig2}}
\end{figure}

In order to verify the newly developed FAM codes, benchmark tests have been performed together with the conventional RPA code.
The transition strengths of ISGMR in $^{208}$Pb calculated by m-FAM are compared with the conventional RPA results in Fig.~\ref{Fig2}, where all of the common numerical parameters are the same, including $R=25$~fm, $dr=0.1$~fm, single-particle energy truncation $[-M, M+200~\mbox{MeV}]$, and $\Gamma=1$~MeV.
One can barely distinguish these two lines in the figure.

Although we do not show one by one, we have also performed the conventional RPA calculations for the cases without Dirac sea or without rearrangement terms discussed below.
It is found that all of these results are identical to those by the i-FAM and m-FAM calculations.
This demonstrates the feasibility and accuracy of the present FAM codes.

\subsection{Box size}

\begin{figure}
\centering
\includegraphics[width=8cm]{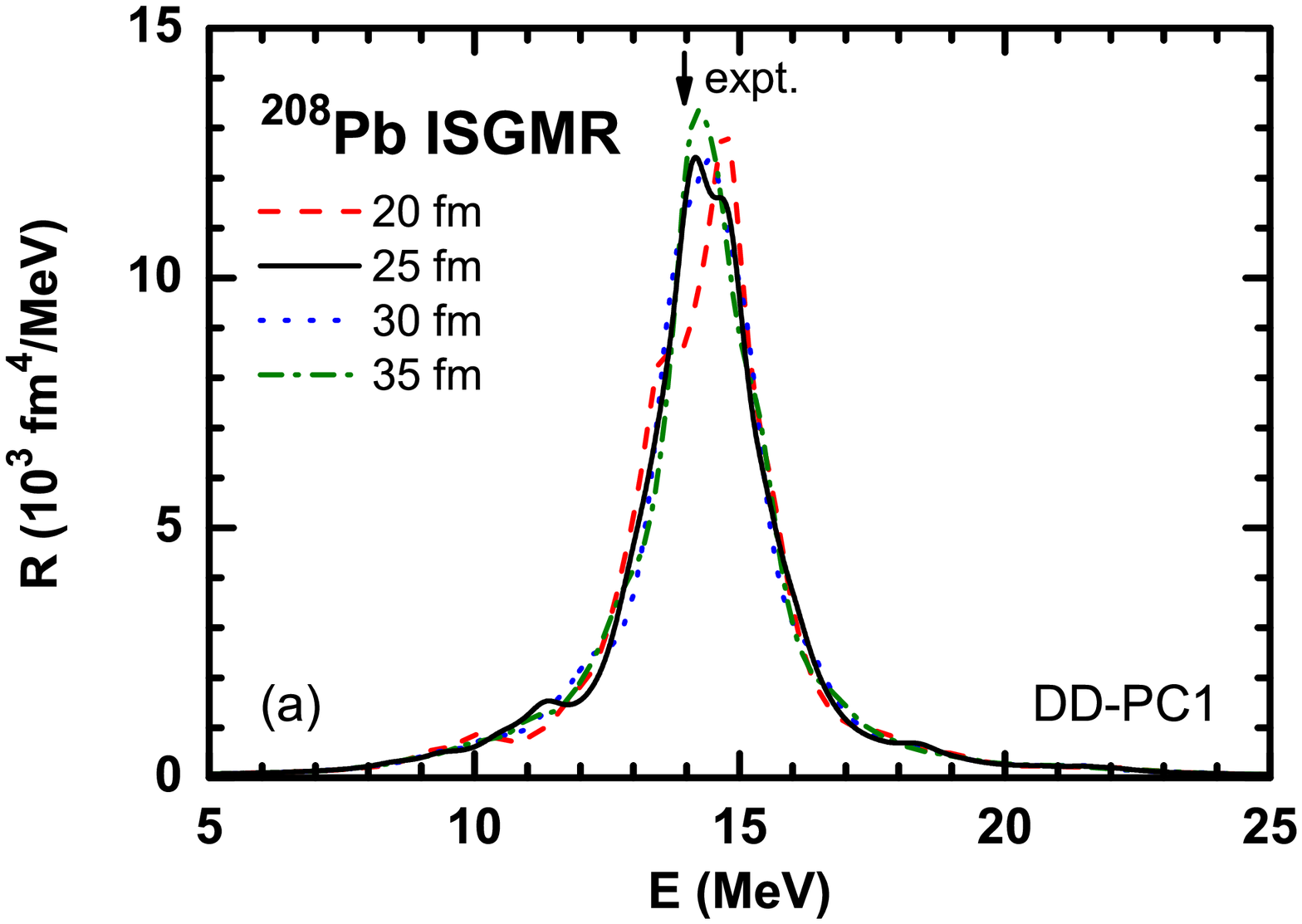}\\
\includegraphics[width=8cm]{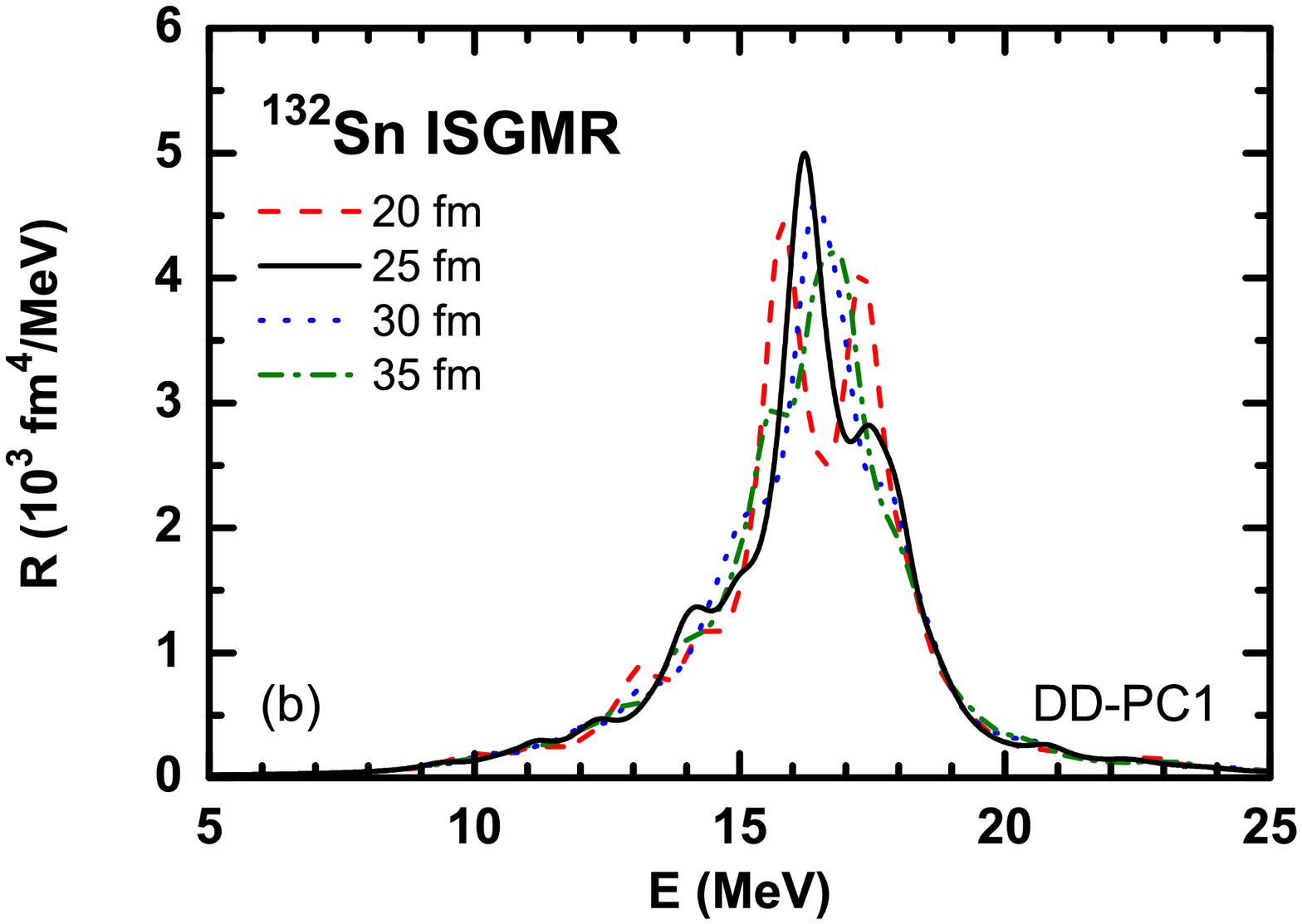}
\caption{(Color online) ISGMR in (a) $^{208}$Pb and (b) $^{132}$Sn calculated by m-FAM with different box size $R$.
The results calculated with $R=20, 25, 30, 35$~fm are shown with the dashed, solid, dotted, and dash-dotted lines, respectively.
The experimental centroid energy in $^{208}$Pb \cite{Youngblood2004} is denoted by the arrow.
    \label{Fig3}}
\end{figure}

In Fig.~\ref{Fig3}, the transition strengths of ISGMR in $^{208}$Pb and $^{132}$Sn calculated by m-FAM with box sizes $R=20, 25, 30, 35$~fm are shown with the dashed, solid, dotted, and dash-dotted lines, respectively.
It is shown that the detailed shapes of the resonances change with $R$ to some extents.
Nevertheless, one of the most important properties, the centroid energy $m_1/m_0$, does not depend on $R$ up to the digit of $0.01$~MeV.
Integrating the excitation energy from $5$ to $25$~MeV, the centroid energies in $^{208}$Pb and $^{132}$Sn are $14.33$ and $16.28$~MeV, respectively.
The experimental data in $^{208}$Pb, $m_1/m_0 = 13.96\pm0.20$~MeV \cite{Youngblood2004}, can be well reproduced.
In the following calculations, the box size $R=25$~fm and the mesh size $dr=0.1$~fm are used.

\subsection{Effects of the Dirac sea}

Comparing with the non-relativistic FAM, it is interesting to investigate the effects of the Dirac sea in the relativistic FAM calculations, in particular, for those using the coordinate-space representation.

\begin{figure}
\centering
\includegraphics[width=8cm]{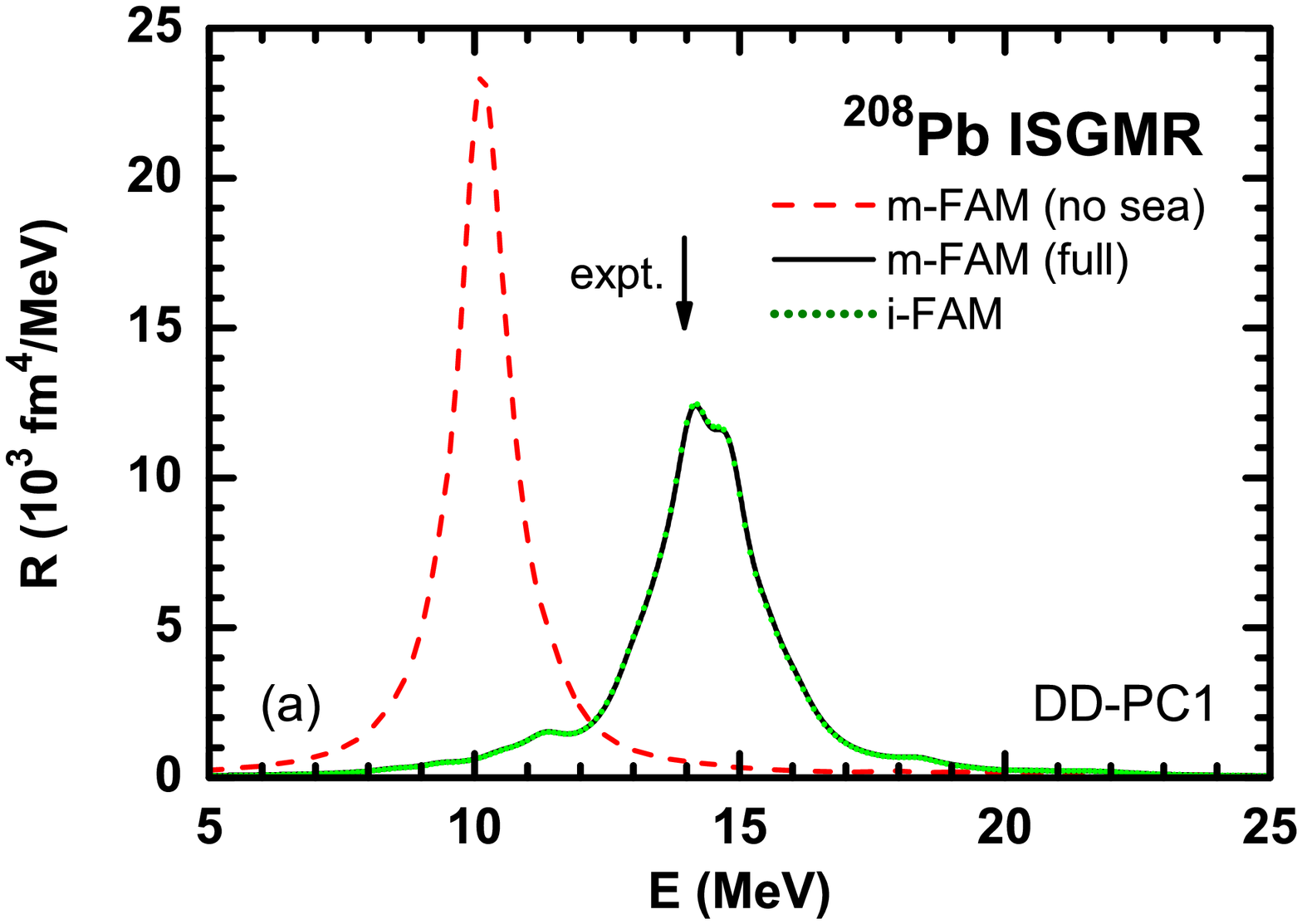}\\
\includegraphics[width=8cm]{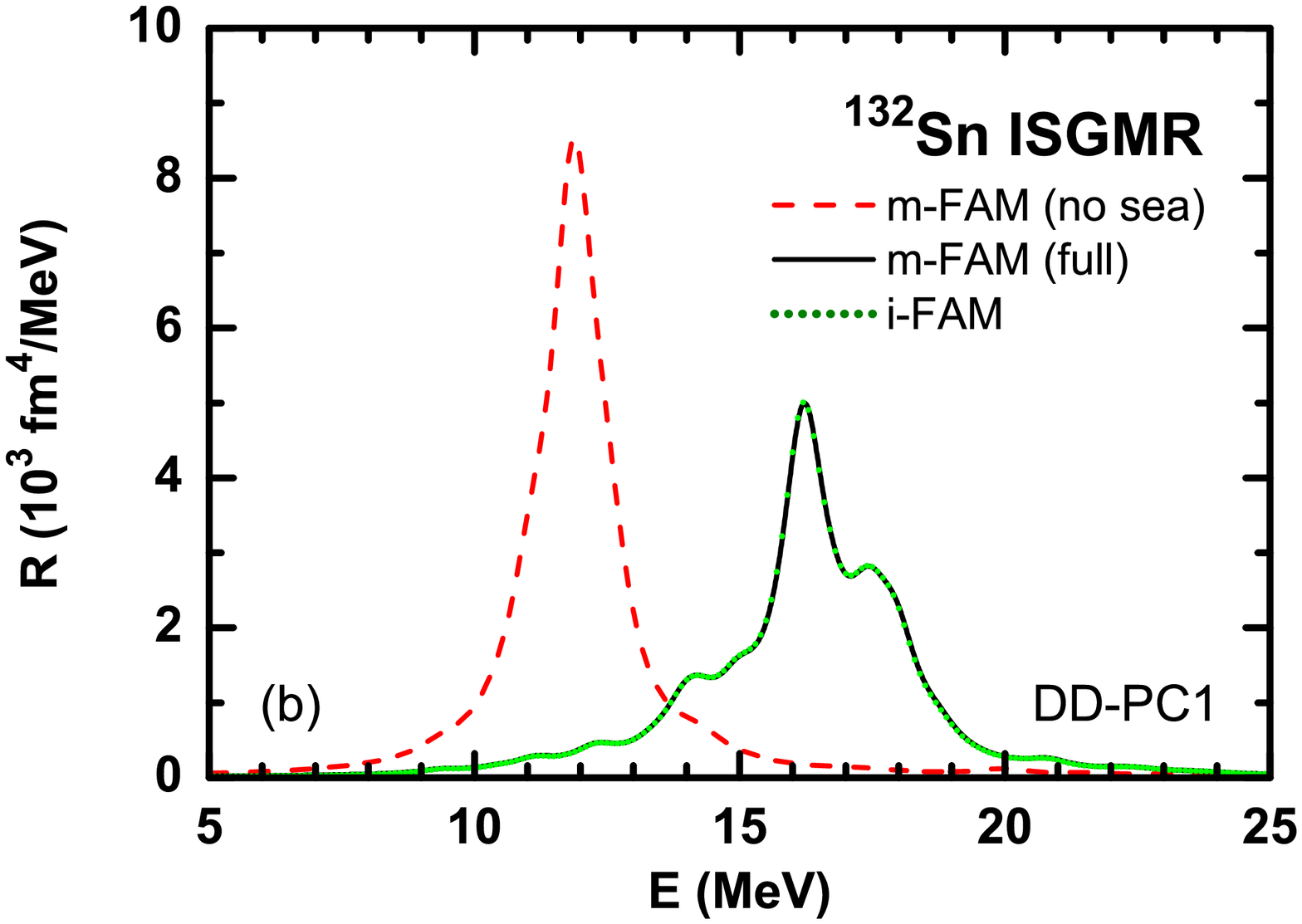}
\caption{(Color online) ISGMR in (a) $^{208}$Pb and (b) $^{132}$Sn calculated by i-FAM and m-FAM.
The i-FAM results are shown with the dotted symbols, while the m-FAM results calculated with and without the Dirac sea are shown with the solid and dashed lines, respectively.
The experimental centroid energy in $^{208}$Pb \cite{Youngblood2004} is denoted by the arrow.
    \label{Fig4}}
\end{figure}

The effects of the Dirac sea can be explicitly identified in the m-FAM calculations.
In Fig.~\ref{Fig4}, the transition strengths of ISGMR in $^{208}$Pb and $^{132}$Sn calculated with and without the Dirac sea are compared.
The results including the configurations formed from the occupied states in the Fermi sea and unoccupied negative-energy states in the Dirac sea are shown with the solid line, while the results excluding these configurations are shown with the dashed line.
It is found that the Dirac sea effects on the centroid energies $m_1/m_0$ of ISGMR in $^{208}$Pb and $^{132}$Sn are as much as $4.00$ and $4.26$~MeV, respectively.
This substantial influence is due to the strong couplings between the Fermi sea and Dirac sea in the scalar channel \cite{Ring2001}.
The experimental data \cite{Youngblood2004} is reproduced only when the Dirac sea is taken into account.

In the coordinate-space representation as in i-FAM, one can identify no other single-particle eigenstates but only the occupied states in the Fermi sea.
Just from the mathematical point of view, the coordinate space should generate another complete set of basis for particle states.
In Fig.~\ref{Fig4}, we also plot the corresponding i-FAM results with the dotted symbols by taking the energy spacing $\Delta E=0.1$~MeV.
It can be clearly seen that the i-FAM results are exactly on top of the m-FAM results that include the Dirac sea.
This confirms that these two different sets of basis are both complete and these two methods are equivalent.
This also demonstrates that the existence of Dirac sea does not introduce additional difficulties for the present iterative method in the relativistic scheme, while the only price to pay is that the total dimension of the i-FAM equations (\ref{eq:iFAM}) is now as twice as the non-relativistic counterpart.

\subsection{Effects of the rearrangement terms}

It is tedious to calculate the contributions of the rearrangement terms in $V_{ph}$ to the RPA matrix elements in the conventional calculations.
From Eq.~(\ref{eq:Vph}), one can see that, for one normal term in each channel, there are up to 3 rearrangement terms accompanied.
In fact, in the meson-exchange picture, this number increases to 6 as shown in Ref.~\cite{Niksic2002}.
Even worse, in the RPA based on the density-dependent relativistic Hartree-Fock theory, the number of rearrangement terms accompanied can be $\sim10^2$ as a result of an additional summation over the occupied orbitals due to the non-locality of the self-energies \cite{Liang2010thesis}.

In contrast, as illustrated in Sections~\ref{sec:iFAM} and \ref{sec:mFAM}, the effects of the rearrangement terms can be simply taken into account in FAM by re-calculating the coupling strengths $\alpha$ in Eq.~(\ref{eq:Sigma}) and their derivatives $\partial\alpha/\partial\rho_b$ in Eq.~(\ref{eq:SigmaR}) with Eq.~(\ref{eq:alpha}) for each given set of $\lc\psi'\rl$ and $\lr\psi\rc$.
The numerical cost of such a step is totally negligible, thus this method is extremely efficient.

\begin{figure}
\centering
\includegraphics[width=8cm]{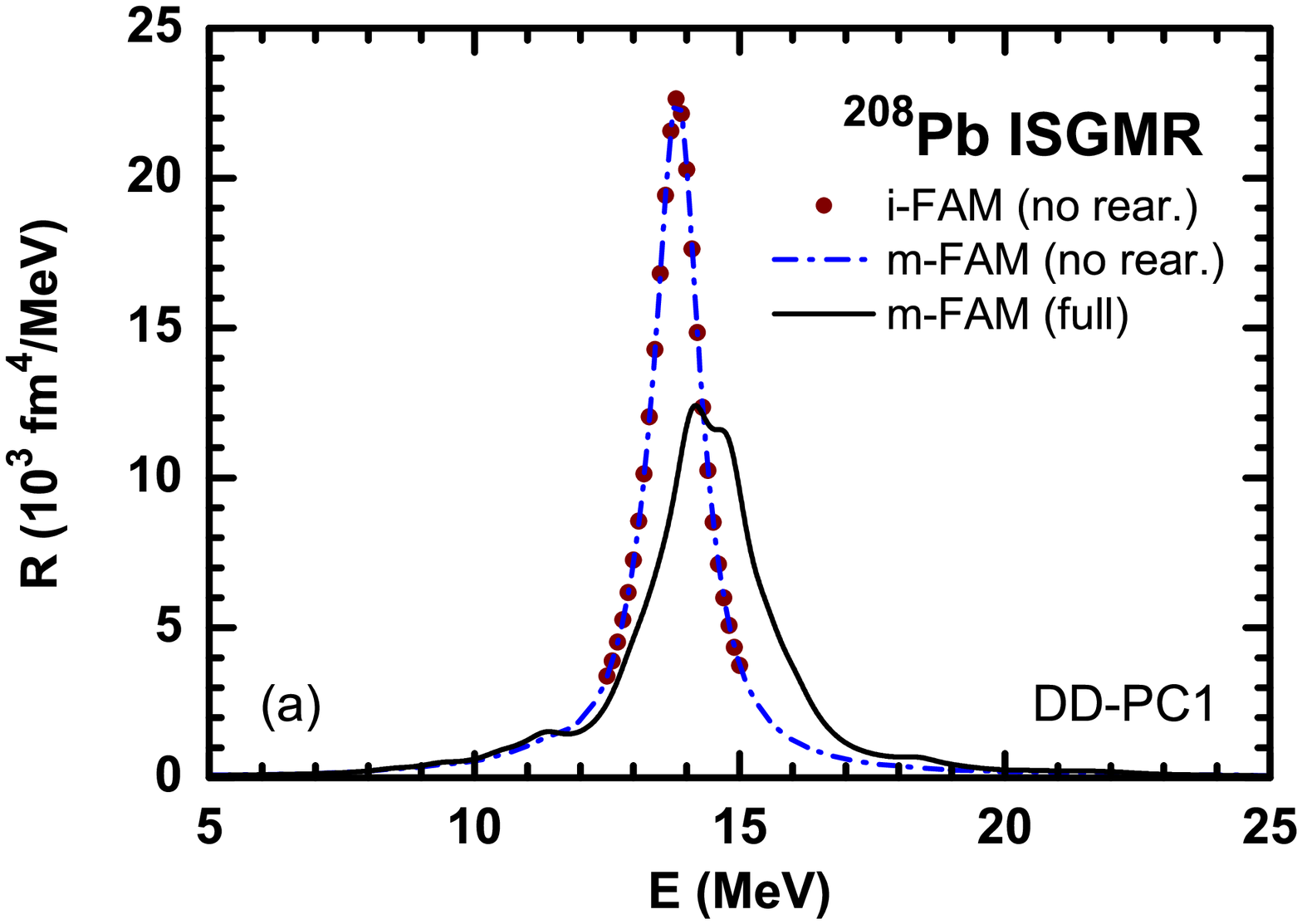}\\
\includegraphics[width=8cm]{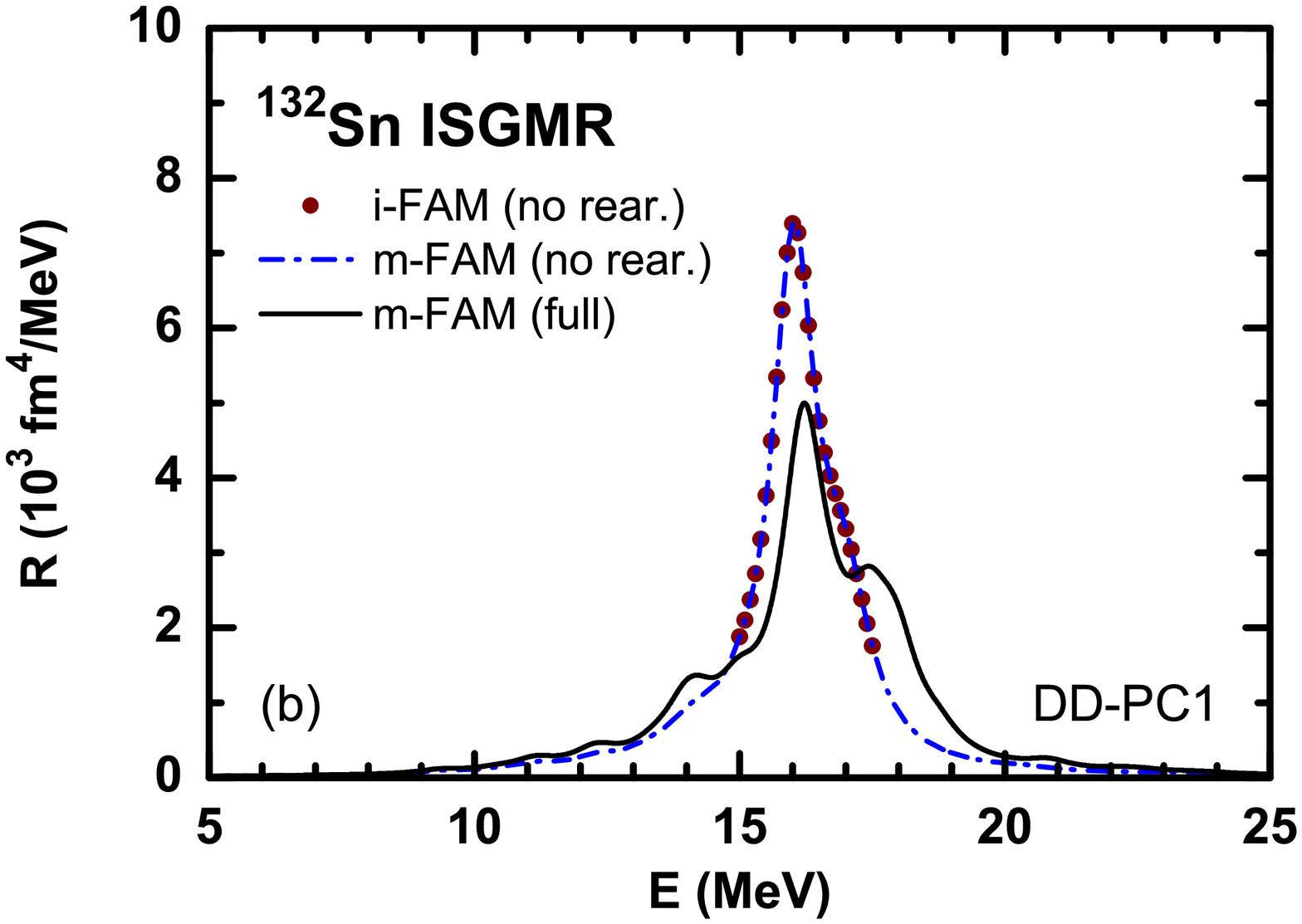}
\caption{(Color online) ISGMR in (a) $^{208}$Pb and (b) $^{132}$Sn calculated by i-FAM and m-FAM.
The i-FAM results without the rearrangement terms are shown with the dotted symbols, while the m-FAM results calculated with and without the rearrangement terms are shown with the solid and dash-dotted lines, respectively.
    \label{Fig5}}
\end{figure}

In Fig.~\ref{Fig5}, the transition strengths of ISGMR in $^{208}$Pb and $^{132}$Sn calculated by m-FAM with and without the rearrangement terms are shown with the solid and dash-dotted lines, respectively.
Around the main-peak region, by taking $\Delta E=0.1$~MeV, the i-FAM results calculated without the rearrangement terms are also shown with the dotted symbols for comparison.
The equivalency of these two finite amplitude methods is illustrated once more, since the rearrangement terms can be switched on or off in the same way.
Quantitatively, it is found that the rearrangement effects on the centroid energies $m_1/m_0$ of ISGMR in $^{208}$Pb and $^{132}$Sn are $0.53$ and $ 0.26$~MeV, respectively, which are also substantial.

%%==================Summary and Perspectives==============================
\section{Summary}\label{sec:SP}

Based on the spherical density-dependent point-coupling RMF theory, the self-consistent relativistic RPA approaches have been established by using the finite amplitude method, where the i-FAM and m-FAM schemes are employed, respectively.

For the FAM coding and calculations, the time-odd components of the functional, i.e., the nucleon currents and the space-component of the Coulomb field, must be kept explicitly.
In the present covariant density functional, these time-odd components have the same coupling strengths as the corresponding time-even components due to the Lorentz symmetry.
This makes the extension of FAM straightforward.
Another key point for the FAM coding is the difference between the single-particle wave functions and their Hermitian conjugates.
The formulas related to these key points are shown in Sec.~\ref{sec:TF} in details.

By taking the ISGMR in $^{208}$Pb and $^{132}$Sn as examples, the newly developed methods are verified by the conventional RPA calculations.
It is also found that although the detailed shapes of the resonances depend on the box size $R$ to some extents, the calculated centroid energies $m_1/m_0$ are precise up to $0.01$~MeV.
The experimental data in $^{208}$Pb is well reproduced.

For the effects of the Dirac sea, it is confirmed that the \textit{ph} configurations concerning the particle states in the Dirac sea must be included explicitly in the m-FAM scheme.
On the other hand, such effects can be automatically taken into account in the coordinate-space representation as in the i-FAM scheme, because the coordinate space, $\sum_{\mathbf{r}}\lr\mathbf{r}\rc\lc\mathbf{r}\rl-\sum_j\lr\phi_j\rc\lc\phi_j\rl$, provides an equivalent complete set of basis for particle states.
For the rearrangement terms, instead of being calculated term by term in the conventional RPA, they can be implicitly calculated without extra computational costs in both i-FAM and m-FAM schemes.
One simply needs to re-calculate the coupling strengths $\alpha$ and their derivatives $\partial\alpha/\partial\rho_b$ for each given set of $\lc\psi'\rl$ and $\lr\psi\rc$.

In conclusion, the feasibility of the FAM for the covariant density functionals has been demonstrated, and the advantages on treating the Dirac sea and rearrangement terms in the relativistic RPA have been presented.
This opens a new door for developing the self-consistent relativistic RPA for deformed nuclei.

%%==================Acknowledgments==============================
\section*{Acknowledgments}

The authors are grateful to Dr. Paolo Avogadro for the helpful discussions on the GCR solver.
This work is partly supported by
the Grant-in-Aid for JSPS Fellows under Grant No. 24-02201,
the JSPS KAKENHI under Grants No. 20105003 and No. 21340073,
the Major State 973 Program 2013CB834400,
the National Natural Science Foundation of China under Grants No. 10975008, No. 11105006, No. 11175002, and No. 11205004,
the Research Fund for the Doctoral Program of Higher Education under Grant No. 20110001110087,
the 211 Project of Anhui University under Grant No. 02303319-33190135.

%\bibliography{FAM}

%\end{CJK*}
\end{document}